\documentstyle[twocolumn,aps,epsf]{revtex}
\begin{document}
\newcommand{\beq}{\begin{equation}}
\newcommand{\eeq}{\end{equation}}

\title{DELINEATION OF THE NATIVE BASIN IN CONTINUUM MODELS OF PROTEINS}

\author{Mai Suan Li and Marek Cieplak}

\address{Institute of Physics, Polish Academy of Sciences, 
Al. Lotnikow 32/46, 02-668 Warsaw, Poland }

\address{
\centering{
\medskip\em
{}~\\
\begin{minipage}{14cm}
We propose two approaches for determining 
the native basins in off-lattice models of proteins.
The first of them is based on exploring the saddle points on
selected trajectories emerging from the native state. In the second
approach, the basin size
can be determined by monitoring random distortions in the shape of
the protein around the native state. Both techniques yield the similar
results. As a byproduct, a simple method to determine the folding
temperature is obtained.
{}~\\
{}~
\end{minipage}
}}

\maketitle

Chains of beads on the cubic or square lattices, 
with some effective interactions
between the beads, often serve as simple models of proteins
(see for instance ref.\cite{1}).
A more realistic modelling, however, requires considering
off-lattice systems.
Simple off-lattice heteroplymers have been
discussed recently by  Iori, Marinari, and Parisi\cite{2},
Irback et al.\cite{3},
Klimov and Thirumalai\cite{4}, and by the present authors\cite{5}.
The purpose to use such models  is to understand the basic mechanism of
folding to the native state. In lattice models, the native state
is usually non-degenerate  and it coincides
with the ground state of the system. 
Delineating boundaries of the native basin in off-lattice systems,
however, is difficult, especially  when the number of degrees 
of freedom is large , yet it is essential for studies of
almost all equilibrium and dynamical properties of proteins.
For instance, stability of a protein is determined by estimating
the equilibrium probability to stay in the native basin: 
the temperature at which this
probability is $\frac{1}{2}$ defines the folding temperature, $T_f$.
The native basin consists of the native state and its
immediate neighborhood, as shown schematically in Figure 1, and it
should not be confused with the whole folding funnel. The latter involves a 
much larger set of conformations which are linked kinetically to the 
native state.

In most studies, such as in Ref.\cite{3,4}, the size of a basin
is declared  by adopting a reasonable but {\em ad hoc}
cutoff bound. Systematic approaches, however, are needed
and will be presented here.
The task of delineating of the native basin is facilitated 
by introducing the concept of a distance between two conformations
$a$ and $b$, $\delta _{ab}$. The distance should be defined in a way
that excludes effects of an overall translation or rotation.
There are two definitions of $\delta _{ab}$, for a sequence
of $N$ monomers, that we shall use.
The first one is\cite{2,6}:
\begin{equation}
\delta_{ab}^2 \; \; = \; \; $min$ \frac{1}{N} \sum_{i=1}^N 
| \vec{r}_i^a - \vec{r}_i^b |^2 \; \; ,
\end{equation}
where $\vec{r}_i^{a}$ denotes the position of monomer $i$ in 
conformation $a$ and the minimization is  performed over translations,
rotations and reflections.
In practice, we put chain $a$ over chain $b$ by
overlapping the two centers of mass, and then we find the optimal rotation
of $b$ which minimizes $\delta_{ab}$. 

\begin{figure}
\epsfxsize=3.2in
\centerline{\epsffile{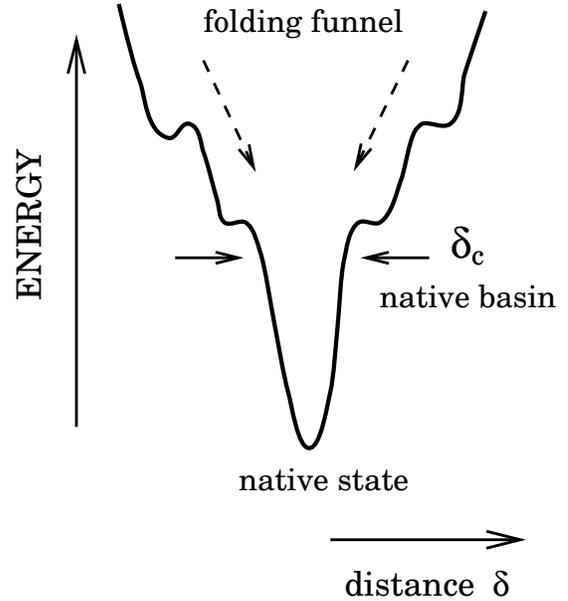}}
\caption{A schematic plot of the energy in the vicinity
of the native state. $\delta _c$ denotes a characteristic
size of the basin.}
\end{figure}

The second is
\begin{equation}
\delta'^2_{ab} \; \; = \; \; \frac{1}{N^2-3N+2} \sum_{i \neq j,j\pm 1} 
(| \vec{r}_i^a - \vec{r}_j^a | - | \vec{r}_i^b - \vec{r}_j^b |)^2 \; \; .
\end{equation}

\noindent
The first definition is closer to an intuitive understanding of
the distance between shapes whereas the second is easier to compute,
especially in three-dimensional situations. Nevertheless both are
expected to be physically equivalent.

\begin{figure}
\epsfxsize=3.2in
\centerline{\epsffile{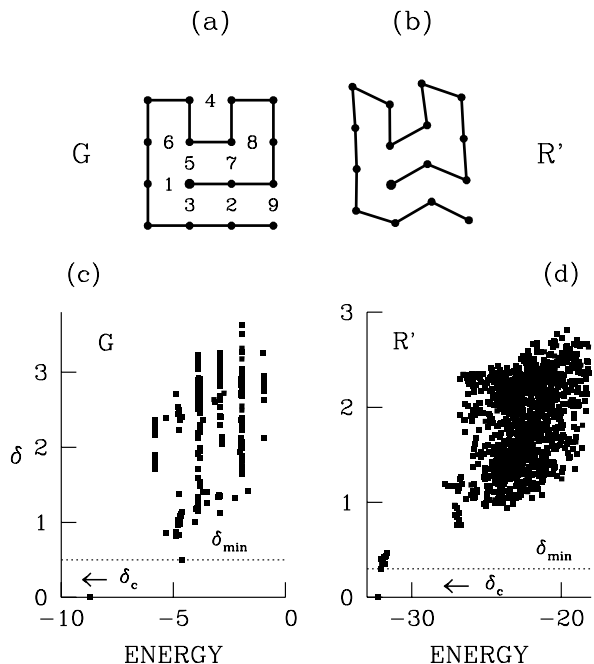}}
\caption{
(a) The target  native conformation for sequence $G$.
Its energy is equal to -8.716 in units of $C$. The figure shows
also the assignment of the couplings $A_{ij}$ used to construct sequence $R'$.
The numbers indicate the relative strengths
of the contacts. (b) The native conformation of sequence $R'$.
(c) The distances $\delta$ from local minima to the native
state, for sequence $G$, shown as a function of energy of the minima.
The dotted line corresponds to the minimal distance
$\delta_{min}$. The horizontal arrow indicates the size of the
basin boundary, as obtained form studies of the shape distortions.
(d) The same as in c) but for sequence $R'$.}
\end{figure}

In this paper, we develop two techniques for determining the basin of
the native state. In the first approach, we generate an 
image of the phase space by sampling it by low energy 
trajectories that start from the native state and continue by
displacing the monomers in a way that
increases the distance away from the native state while
preserving the connectedness of the chain.
For each trajectory, a dependence of the potential
energy on the distance away from the native state is 
obtained and locations of the saddle points are determined.
The average distance to a first encountered saddle  point,
$<\delta_s>$, may be considered as
characterizing the size of the basin.

In the second technique, which we find to be more statistically
reliable and more automatic in its implementation, we adopt a
variant of de Gennes's idea of the "ant in a labirynth"\cite{7}.
Specifically, we
characterize the geometry of the native basin 
by monitoring random shape distortions of the heteropolymer in the basin. 
The distortions
are induced by diffusive-like displacements of individual beads.
The method of the
shape distortion is implemented by first placing the polymer in an
initial conformation, $a(0)$, which usually coincides with the native
state. Subsequently, one performs random displacements
of individual beads in the chain, through a Monte Carlo routine,
and the conformation at time $t$ acquires a shape $a(t)$. 
The process is characterized by determining the evolution of a mean
square distance, $<\delta ^2>_t\;=\;<\delta ^2 _{a(0)a(t)}>$ between
$a(t)$ and $a(0)$. The focus is on short time behavior and the average
is over many trajectories that start from the same $a(0)$.
The characteristic size of the basin, $\delta _c$ is obtained
by studying features in $ < \delta ^2 >_t$ 
as described later.
In order
to scale the walls of the basin one needs to make the polymer
'crawl' up these walls without involving any kinetic energy.
Thus the process does not correspond to any 'real' evolution in time,
as defined e.g. through the molecular dynamics.
As opposed to the first technique, the fluctuations in the
shape of the polymer are understood to be due to coupling to a heat bath.

\noindent
{\bf Models}

In order to illustrate the two techniques,
we consider a two-dimensional version
of the model introduced by Iori, Marinari and Parisi\cite{2}.
The Hamiltonian is given by
\begin{equation}
H\; \; = \; \; \sum_{i \neq j} \{ k (d_{i,j} - d_0)^2 \delta_{i,j+1}
+ 4 [ \frac{C}{d_{i,j}^{12}} - \frac{A_{ij}}{d_{i,j}^6} ] \} \; \; ,
\end{equation}
where $i$ and $j$ range from 1 to the number of beads,
$N$, which in our model is equal to 16. The distance between the beads,
$d_{i,j}$ is defined as $ |\vec{r}_{i}-\vec{r}_{j}|$, where
$\vec{r}_i$ denotes the position of bead $i$, and is measured
in units of the standard Lennard-Jones length parameter $\sigma$.
The harmonic term
in the Hamiltonian
couples the adjacent beads along the chain.  The remaining terms
represent the Lennard-Jones potential.
In \cite{2} $A_{ij}$ is chosen as $A_{ij} = A_0 + \sqrt{\epsilon}\eta_{ij}$,
where $A_0$ is constant,
$\eta_{ij}$'s are Gaussian variables with zero mean and unit variance;
$\epsilon$ controls the strength of the quenched disorder.
The case of $\eta_{ij}=0$ and $A_0= C$ would correspond to a
homopolymer. Our choice for the values of $A_{ij}$ is that
all of the $A_{ij}$'s are positive, which corresponds to attraction.
We measure the energy in units of $C$ and
consider $k$ to be equal to 25 in units of $C/\sigma ^2$. 
Smaller values of $k$ may violate
the self-avoidence of the chain\cite{5}.

\begin{figure}
\epsfxsize=3.2in
\centerline{\epsffile{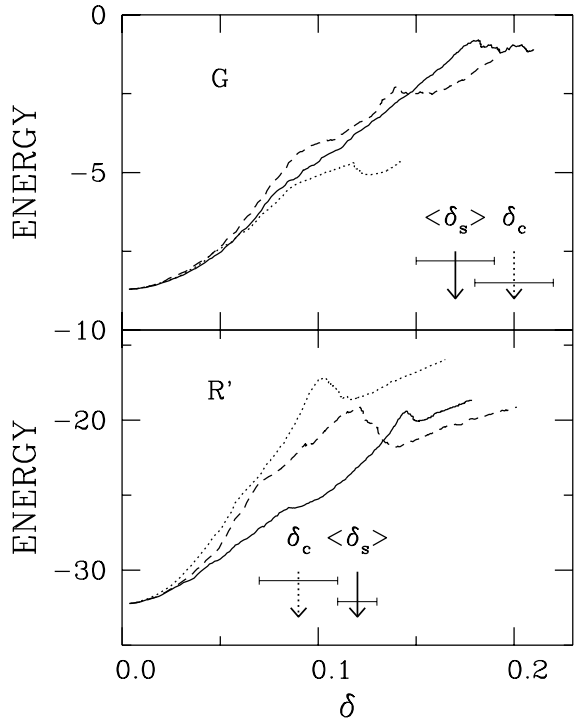}}
\caption{Typical trajectories from the native state generated by the first
method. The solid and dotted arrows denote the basin sizes $<\delta_s>$
and $\delta_c$ obtained by the first and second approaches respectively.}
\end{figure}

We focus on two 16-monomer sequences in two dimensions
which were characterized in detail 
previously\cite{5}. One of them, $G$, is a good folder and the other, $R'$,
is a bad folder.
We used two criteria for the quality of folding: 1) based on the evaluation
of the specific heat and structural susceptibitity \cite{8}
and 2) based on the location of the folding temperature,
$T_f$, relative to the temperature at which the onset of the glassy 
effects takes place \cite{9}.
$G$ has been designed as an off-lattice Go-like
sequence\cite{10} and
the $A_{ij}$'s are taken to be 1 for
native contacts and 0 for non-native ones. The target conformation is on
lattice and is shown in Figure 1a.
This target is the same as for the lattice sequence $R$ of
Ref.\cite{11,12}.
The ground state of $G$ has approximately the shape of the target - note
that the interaction between two beads forming a contact also slightly
affects other beads.
The bad folder $R'$ is constructed following
the rank-ordering procedure  which is an off-lattice analog of what
was done in references\cite{11,12}.
We choose the equlibrium interbead distance, $d_0$ of
$2^{1/6}$ and $1.16$ (the latter value is determined by the
average of $A_{ij}$ over all couplings, see Ref.\cite{5})
for $G$ and $R'$ respectively. 
The target
conformation is chosen initially to be the same as in Figure 1a.
The most strongly attractive Lennard-Jones interactions
are assigned to the nine native contacts.
They are enhanced by making the corresponding $A_{ij}$ bigger than one.
The remaining couplings have $A_{ij}$'s which are smaller than 1.
Rank ordering of the contacts generates good folders among
lattice models. In the off-lattice models, however, the non-native 
Lennard-Jones interactions,
due to their long range nature, overconstrain the system and frustrate
it leading to a deterioration of folding properties. 
The values of $A_{ij}$ for $R'$ are listed in Ref.\cite{5}.
The resulting  native state of $R'$
is shown in Figure 1b.

Figures 2c and 2d show the distributions of distances $\delta$ for
local energy minima of sequences $G$ and $R'$ respectively. The native
states and the local energy minima have been obtained by
multiple quenches from random conformations.
Note that the sequence $G$ has a much fewer number of the local
energy minima compared to sequence $R'$ and the energy gap
to the first excited local minimum is significantly larger.
In each case, there is one minimum which has the closest geometrical
distance, $\delta _{min}$ to the native state. $\delta _{min}$ is thus
an upper bound for the size of the native basin 
(0.5 for $G$ and 0.3 for $R'$).

\noindent
{\bf Trajectories in the energy landscape}

The first technique is implemented by creating the
trajectories from the native state at $T=0$ in a stochastic way.
The bead positions are displaced randomly and 
the moves are accepted if they
increase the geometrical distance
to the native state and keep the distance between nearest beads to be
in the interval $1 < d_{i,i+1} <1.1d_0$.  
In addition one has to keep the interaction energy for any pair 
of monomers in sequence 
$R'$ to be negative. For $G$, however, the non-native pairs interact
only repulsively.
In order to minimize the repulsion as much as possible we explore 
only those trajectories which keep the distances between the
monomers of non-native contacts sufficiently large. We choose the minimal
distance between the
beads of non-native contacts, $d_{m}$, to be $d_{m}=1.5$ (the choice
of 1.6 for $d_m$ yields similar results) and reject trajectories which
lead to a monotonic increase in energy even after entering into
the region of overall positive energies.
Smaller values of $d_{m}$ usually 
lead to trajectories with unreasonably large positive energies. 
Substantially larger values of $d_{m}$
generate short trajectories which terminate on a conformation
which does not allow for a further increase in the distance.

Figure 3 shows typical trajectories for $G$ and $R'$. For each trajectory, 
we define the postion of the saddle point $\delta_s$. 
This point shows as a local maximum on the energy vs. $\delta$ curve.
The average value,
$<\delta_s>$,  defines the basin size. Averaging over 50 trajectories
we obtain $<\delta_s> = 0.17 \pm 0.02$ and $ 0.12\pm 0.01$
for sequences $G$ and $R'$ respectively.

It should be noted that the technique described here is simple
conceptually and easy to implement for a small number of trajectories
but its resulting statistics on the basin size are 
inherently poor since there is no convenient
control over the choice of important trajectories. Following
patterns in the force field deterministically is a possible
improvement but another stochastic approach, described below,
combines simplicity with reliability of the results.

\noindent
{\bf Fluctuations in the shape of the polymer}

In order to compute $<\delta^2>_t$ we update the monomer positions
randomly within circles of radius of 0.01 (the choice of 0.02
yield similar results).  We assume that the system is in contact
with a heat bath corresponding to temperature $T$ which
provides a controlling device.
Figure 4 shows the dependence of  $<\delta^2>_t$ on $t$ for sequence $G$.
We observe that there are, in general, three regimes of
behavior of $<\delta ^2>_t$.  For sequence G, all of the three regimes
appear below $T_c$=0.19 and are shown, in Figure 4, for $T$=0.05 and 0.15.

In the first regime, I, corresponding
to very short time scales during which merely several percents
of a linear size of the basin are covered, one has a power law
\begin{equation}
<\delta^2>_t \; \; \sim \; \; t^{\nu _0} \; \; .
\end{equation}
Pliszka and Marinari\cite{13} have demonstrated 
that $\nu _0$ 
is sensitive to the details of the Hamiltonian.
We find that $\nu _0$ is equal to $0.96 \pm 0.02$ and $0.94 \pm 0.02$
for sequences G and R' respectively. Both values are close to 1
and stay the same for all $T$'s (even for $T$ up to 50) which suggests
a simple diffusive behavior.

In the second regime, II, the plot of $<\delta ^2>_t$ vs. $t$
acquires a $T$-dependence. 
Sommelius\cite{6} has focused on this regime and has 
postulated that the behavior here appears to
follow a power law, at least approximately. The corresponding
exponent $\nu$ is then found to depend on $T$  in a way that relates
to characteristic temperatures of the system.
We have found, however, that this power law is not robust -- the
effective exponent depends on the size of the steps in which one 
implements distortions. More importantly, the spatial extent of this
regime is necessarily limited, and it would probably remain so even
for very long heteropolymers.

\begin{figure}
\epsfxsize=3.2in
\centerline{\epsffile{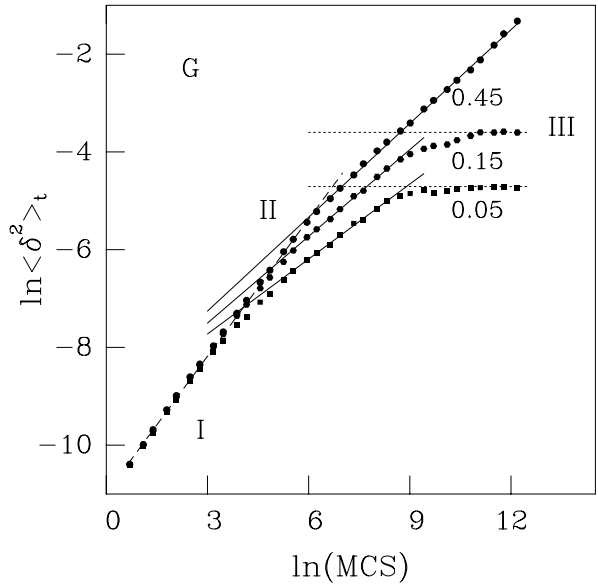}}
\caption{
The dependence of the square of the distance to the native state on
the Monte Carlo time for $G$ for several values of the temperature
which are shown next to the curves.The dashed, solid and dotted
correspond to regimes I, II, and III respectively.
The results are averaged over 400 trajectories. The error bars are
smaller than the symbol sizes. The optimal rotation, required in
Eq. (1), is picked from 1000 discrete values into which
the full 360$^o$ angle is divided.}
\end{figure}

In the third regime, III, observed only below a critical temperature
$T_c$, $<\delta ^2>_t$ saturates at a constant value
as explained in detail in Ref. \cite{5}. Above $T_c$, the
shape distortion ceases to be confined to the native basin and
the type II behavior continues to take place,
as illustrated by the data points corresponding
to $T=0.45$ in Figure 4.  The
limiting saturation value of $\sqrt{<\delta ^2>}$ at $T_c$ defines
a characteristic basin size, $\delta _c$, that can be used, e.g.,
when deciding if a folding took place if the system started
to evolve from and unfolded state. Naturally, $T_c$ is a measure
of the folding temperature $T_f$ which characterizes thermodynamic
stability of the system.

Figure 5 switches from the logarithmic scale of Figure 4 to the linear 
scale  in wich the transition between the staturation and lack
of saturation in $<\delta ^2>_t$ shows in a more convincing way. 
Figure 5  compares the
time dependence of $<\delta ^2>_t$  to that of $<\delta '^2>_t$
for $G$ at $T_c$ and 
demonstrates that the actual choice of the definition
of the distance has small effect on the results.
For sequence $G$ we have
$\delta_c = 0.2 \pm 0.02$ (the second definition of the distance yields
$\delta'_{c}=0.17 \pm 0.02$)
For sequence $R'$ we find that
$T_c\approx 0.09$ and $\delta_c = 0.09 \pm 0.02$. 
Within the error bars, $\delta_c$ is close to $<\delta_s>$ defined
by the first approach which is also indicated in Figure 3.
For both sequences, 
the values of $\delta _c$ and $<\delta_s>$ are smaller than $\delta _{min}$.
Thus {\em the saturation value of the distance to the 
native state at $T=T_c$ may indeed serve as 
a measure of size of the native basin $\delta_c$},
i.e $\delta_c =[<\delta^2>_{sat}(T=T_c)]^{1/2}$ determines the
true boundary of the native basin.

\begin{figure}
\epsfxsize=3.2in
\centerline{\epsffile{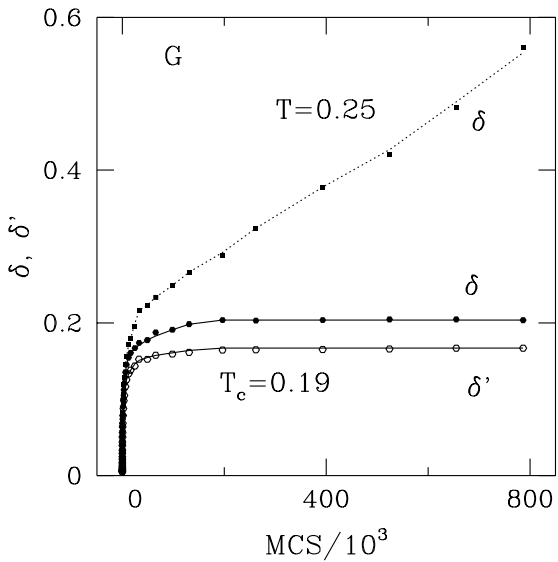}}
\caption{
The time dependence of $\delta = <\delta ^2>_t^{1/2}$
(black hexagons) and $\delta ' =<\delta '^2>_t^{1/2}$
(open hexagons) for sequence $G$
at the critical temperature $T_c$=0.19. For
$T=0.25$, $\delta $ (black squares) does not saturate below the
the distance equal to the minimal distance, $\delta _{min}=0.5$,
between the native state and the local minimum which is the nearest
geometrically (see Figure 2).}
\end{figure}

The fact that $\delta _c$ for $G$ is 
found to be bigger than for $R'$ indicates bigger stability of $G$ relative
to $R'$. This also correlates well with the higher value of $T_c$ 
which in turn suggests that $T_c$ is a measure of $T_f$ -- the folding
temperature. This interpretation is confirmed by comparing the
values of $T_c$ to $T_f$ obtained by calculating the probability
to be within 
the cutoff distance, $\delta _c$, away 
from the native state\cite{5} and by studying positions of the
peaks in the structural susceptibility\cite{5}. These studies yield
$T_f$ of 0.24 $\pm$ 0.03 and $\approx$0.12 for $G$ and $R'$ respectively.
Both of these values are close to $T_c$ obtained from the shape distortion.
It should be noted that the calculation of $T_c$
by monitoring stochastic shape distortions is significantly less
involving computationally.
We have also used this technique to determine the sizes of basins
of low lying local energy minima.
The corresponding values of $\delta _c$
are found to be smaller than for the native state.


We now consider, following Struglia \cite{14}, the basin of
attraction for random initial conformations which are
subsequently quenched in the steepest descent fashion.
The basin of attraction is defined in terms of
a distance at which the probability to fall onto the native state is
bigger than or equal to $p_c$. 
The corresponding basin size will be denoted by $\delta_f$.
In Ref. \cite{14}, $p_c$ was taken to be equal to $\frac{1}{2}$. 
In our studies, we took
$p_c=1$ and determined the basin of attraction for  sequences $G$ and $R'$
through a standard quenching procedure. We considered 200 trajectories 
and obtained
$\delta_f \approx 0.55$ and $\approx 0.35$ for sequences
$G$ and $R'$ respectively. These values exceed not only our estimate
of $\delta _c$ but also that of
the minimal distance
between the native state and the nearest minimum $\delta_{min}$. 
The values of $\delta _f$ would become even larger for a $p_c$
that was less than 1.
This emphasizes the point that the procedure used by Struglia
probably delineates the folding funnel and not the native basin itself.


In summary, we have explored the native basins of two off-lattice sequences 
by monitoring the shape distortion and by exploring the saddle points
of the trajectories from the native state. We
 have come out with computationally
simple methods to delineate the boundaries of the basins and to estimate the
folding temperature.
The bad and good folders are found to have native basins which are
comparable in size even though the structure of their folding funnels
must be very different.

We thank T. X. Hoang for discussions.
This work was supported by Komitet Badan Naukowych (Poland; Grant number\
2P03B-025-13).

\vspace{0.5cm}



\end{document}